# Моделирование реальных значений КПД высокоэффективных кремниевых солнечных элементов

А.В. Саченко[1], А.И. Шкребтий[2], Р.М. Коркишко[1], В.П. Костылев[1], Н.Р. Кулиш[1], И.О. Соколовский[1]

[1] V. Lashkaryov Institute of Semiconductor Physics, NAS of Ukraine
41, prospect Nauky, 03028 Kyiv, Ukraine

[2]University of Ontario Institute of Technology, Oshawa, ON, Canada

E-mail: sach@isp.kiev.ua

Аннотация

Рассчитаны температурные зависимости КПД $\eta$ высокоэффективных солнечных элементов (СЭ) на основе кремния. Показано, что температурный коэффициент падения величины $\eta$ с ростом температуры тем ниже, чем меньше скорость поверхностной рекомбинации.

Выполнено моделирование эффективности фотопреобразования высокоэффективных СЭ на основе кремния, работающих в полевых (натурных) условиях. Их рабочая температура определялась самосогласованно, путем совместного решения уравнений для фототока, фотонапряжения, а также уравнения баланса потоков энергии. Учтены радиационный и конвекционный механизмы охлаждения. Показано, что рабочая температура СЭ выше температуры окружающей среды даже при очень больших коэффициентах конвекции (~300 Вт/м$^2$·К). Соответственно, эффективность фотопребразования при этом меньше, чем в случае, когда температура СЭ равна температуре окружающей среды.

Получены и обсуждены расчетные зависимости для напряжения разомкнутой цепи и эффективности фотопреобразования высококачественных кремниевых СЭ при концентрированном освещении с учетом реальной температуры СЭ.

Abstract

**Modeling of high-efficiency silicon solar cells in realistic operating conditions**

A.V. Sachenko, A.I. Shkrebtii, R.M. Korkishko, V.P. Kostylyov, N.R. Kulish, I.O. Sokolovskyi

The selfconsistent model for the temperature dependence of photoconversion efficiency $\eta$ for highly efficient silicon solar cells (SCs) is developed. It is demonstrated that effect of the efficiency decrease due to increasing temperature is less pronounced in the SCs with lower surface recombination velocity, thus offering a possibility to improve the cells' performance.

The photoconversion efficiency of the high efficiency silicon solar cells is modeled for the realistic ambient conditions. The SC operating temperature is determined by self-consistently solving the photocurrent, photovoltage, and energy balance equations, considering both radiative and convective cooling mechanisms. The SC temperature is shown to be substantially higher than the ambient temperature even at very high convection coefficients, such as, *e.g.*, 300 W / (m$^2$ · K), used in our examples. The photoconversion efficiency for this case is substantially below the efficiency of thermally stabilized SC, for which the operating temperature is close to the external temperature.

The open-circuit voltage and photoconversion efficiency of the high-quality silicon solar cells under concentrated illumination are also investigated including the tradeoff between SCs heating and cooling processes.

## Введение

Высокоэффективными мы будем называть кремниевые солнечные элементы (СЭ) с КПД $\geq 20$ %. В настоящее время созданы кремниевые СЭ с эффективностью фотопреобразования порядка 25 %. Так, в частности, в работе [1] приведены особенности конструкции и результаты исследования СЭ, на котором была достигнута рекордная эффективность фотопреобразования в кремниевом p-n переходе, равная 25 % в условиях AM 1.5. В последние годы в так называемых HIT (heterojuction with intristic thin layer) элементах с одним анизотипным гетеропереходом также достигнуты значения эффективности фотопреобразования $\eta$, значительно превышающие 20 %. В частности, в работе [2] получена величина $\eta$, равная 24. 7 % в условиях AM 1.5. В указанной работе для создания анизотипного гетероперехода на фронтальной поверхности и изотипного гетероперехода на тыловой поверхности использовался $\alpha - Si:H$, а в качестве основного материала – высококачественный монокристаллический кремний с временем жизни Шокли-Рида-Холла $\tau_{SR}$ порядка одной миллисекунды. Как показано в работе [3], такой КПД удалось достичь благодаря минимизации значений эффективных скоростей поверхностной рекомбинации на освещенной и тыловой поверхностях $S_0$ и $S_d$, получив их суммарное значение менее 10 см/с. В этом определяющую роль сыграли сверхтонкие слои $\alpha - Si:H$, толщиной порядка 10 нм, обеспечившие, с одной стороны, согласование решеток используемых материалов, а, с другой стороны, проявилось пассивирующее действие водорода на плотность пограничных рекомбинационных состояний.

Следует, однако, отметить, что в реальных (натурных) условиях функционирования солнечных элементов (СЭ) величина эффективности фотопреобразования $\eta$ будет отличаться от значения, полученного в условиях AM 1,5. Причин такого отличия несколько. Во-первых, температура окружающей среды $T_0$ в условиях работающего СЭ отличается от значения 298 К, т.е. 25° С, при котором производится сертификация СЭ в условиях AM 1.5. Во-вторых, реальная температура СЭ всегда превышает температуру окружающей среды и определять ее нужно, учитывая механизмы охлаждения, в частности, радиационное и конвекционное охлаждение. Наконец, сами условия освещения зависят от географической широты, от времени года, а также от времени дня [4,5]. Обычно спектральные условия освещения описывают как AM n, где $n = 1/\cos(\theta)$, $\theta$- азимутальный угол между вертикалью и направлением на Солнце. В работах [4,5] условия освещения определялись моделированием поглощения и рассеяния света в атмосфере при произвольных углах падения с учетом следующих механизмов поглощения и рассеяния света: поглощение парами воды, поглощение озоном, поглощение двуокисью азота, поглощение аэрозолями, поглощение смесью остальных газов, релеевское рассеяние.

Для того, чтобы определить рабочую температуру СЭ в реальном случае, кроме уравнений для фототока и фотонапряжения нужно решить уравнение баланса температуры (вернее говоря, уравнение баланса потоков энергии) . Оно имеет следующий вид [6]:

$$P_s(\varepsilon - \eta(T)) = \beta K_T \sigma\left(T^4 - T_0^4\right) + \gamma(T - T_0). \quad (1)$$

Здесь $P_s = \int_{E_1}^{E_2} P(E_{ph}) dE_{ph}$ - плотность мощности солнечного излучения, $P(E_{ph})$ - удельная плотность мощности солнечного излучения для данной энергии фотона, $E_1$ и $E_2$ - граничные энергии фотонов, задающие энергетический интервал солнечного излучения. Для кремниевых СЭ, в которых время жизни Шокли-Рида-Холла, как правило, значительно меньше излучательного времени жизни, величина $\varepsilon \approx 1$. Значение $\beta$ порядка 1. Оно зависит от коэффициента серости, т.е. от близости теплового излучения СЭ к излучению абсолютно черного тела. Величина $K_T$ равна отношению площади излучающей поверхности к площади освещаемой поверхности СЭ. Параметр $\sigma$ - постоянная Стефана – Больцмана. Величина $T = T_0 + \Delta T$, $\Delta T$ - величина превышения полной температуры Т по сравнению со значением $T_0$, $\gamma = \gamma_0 + \delta V_w$ [7] - коэффициент конвекции ($\gamma_0$ - коэффициент конвекции в отсутствие ветра, $V_w$ - скорость ветра, $\delta$ - коэффициент пропорциональности). Следует отметить, что величина коэффициента конвекции кроме скорости ветра зависит от влажности воздуха, от угла между вектором скорости ветра и плоскостью СЭ, а также от барометрического давления.

В настоящей работе мы определим эффективность фотопреобразования высококачественных кремниевых солнечных элементов, положив за основу подход, развитый в [3], а затем решим полученное уравнение для величины $\eta$ совместно с уравнением (1). Это позволит рассчитать теоретически величину эффективности фотопреобразования указанных СЭ в реальных условиях функционирования. Для простоты мы ограничимся рассмотрением случая, когда условия освещения соответствуют спектру для АМ1.5. Полное согласование достигается, как было отмечено выше, конкретизацией условий освещения. Поскольку эта часть задачи была учтена при нахождении эффективности фотопреобразования в работах [4,5], здесь мы остановимся на более подробном анализе той части задачи, которая связана с самосогласованным определением температуры СЭ.

## 1. Постановка задачи

Полученные ниже соотношения для КПД высокоэффективных СЭ на основе кремния справедливы при реализации следующих критериев: 1) длина диффузии $L$ существенно превышает толщину СЭ; 2) концентрация генерированных светом электронно-дырочных пар $\Delta p$ в условиях разомкнутой цепи порядка или больше концентрации основных носителей в базе $p_0$; 3) рекомбинацией в области пространственного заряда в кремнии можно пренебречь по сравнению с рекомбинацией в квазинейтральной области кремния. Как показывает анализ, при выполнении двух первых критериев выполняется и третий.

Поскольку высокоэффективные СЭ производятся из высококачественного монокристаллического кремния со временем жизни Шокли-Рида-Холла порядка или больше 1 мс, то длина диффузии $L$ в них может достигать более одного миллиметра и существенно превышает толщину базы $d$. Т.о. первый критерий в них выполняется. Избыточная концентрация электронно-дырочных пар $\Delta p$ в режиме разомкнутой цепи достигает значений порядка $10^{16}$ см$^{-3}$. Если уровень легирования базы $p_0$ порядка или меньше $\Delta p$, то выполняется и второй критерий. При этом величина напряжения разомкнутой цепи $V_{OC}$ в полупроводнике $p$ - типа определяется таким выражением [3]

$$V_{OC} \cong \frac{kT}{q}\ln\left(\frac{\Delta p}{n_0}\right) + \frac{kT}{q}\ln\left(1+\frac{\Delta p}{p_0}\right). \quad (2)$$

Здесь $p_0$ и $n_0$ - равновесные концентрации основных и неосновных носителей заряда в базе, $k$ - постоянная Больцмана, $q$ - элементарный заряд.

Чем больше отношение $\Delta p / p_0$ по сравнению с 1, тем больше величина напряжения разомкнутой цепи в высокоэффективном кремниевом СЭ по сравнению со случаем, когда $L < d$. Это связано не только с реализацией большого уровня возбуждения ($\Delta p > p_0$), но и с тем, что тыльная поверхность вносит в величину $V_{OC}$ заметный вклад по сравнению с вкладом освещенной поверхности.

Величина $\Delta p$ находится из уравнения баланса генерации-рекомбинации для режима разомкнутой цепи

$$J_{SC}/q = \left[d\left(\tau_{SR}^{-1} + \tau_r^{-1}\right) + S + R_{Auger}\right]\Delta p, \quad (3)$$

где $J_{SC}$ - плотность тока короткого замыкания, $d$ - толщина СЭ, $\tau_{SR}$ - время жизни Шокли-Рида-Холла $\tau_r = \left(A\left(p_0 + \Delta p\right)\right)^{-1}$ - излучательное время жизни $A = 6\cdot 10^{-15}$ см$^3$/с [8] – константа излучательной рекомбинации в кремнии, $S$ - скорость поверхностной рекомбинации,

$$R_{Auger} = C_p\left(p_0 + \Delta p\right)^2 + C_n\left(p_0 + \Delta p\right)\Delta p, \; C_p = 10^{-31}\text{ см}^6\text{/с}, \; C_n = \left(2.8\cdot 10^{-31} + \frac{2.5\cdot 10^{-22}}{\Delta p^{0.5}}\right)\text{ см}^6\text{/с}.$$

Отметим, что уравнение (2) является квадратным уравнением относительно величины $\Delta p$ и его решение имеет такой вид:

$$\Delta p = -\frac{p_0}{2} + \sqrt{\frac{p_0^2}{4} + n_i^2 \exp\left(\frac{qV_{OC}}{kT}\right)}. \qquad (4)$$

Заменив значение $V_{OC}$ на величину приложенного прямого смещения $V$, получаем связь $\Delta p$ с $V$. Это дает возможность сконструировать на основе уравнения вида (3) (не приравнивая правую и левую части) вольт-амперную характеристику для СЭ

$$J(V) = J_{sc} - J_{rec}(V). \qquad (5)$$

Далее из условия максимальной отбираемой мощности $d(VJ(V))/dV = 0$ находится значение $V_m$, а его подстановка в (5) позволяет определить величину $J_m$.

В результате получаем для эффективности фотопреобразования СЭ единичной площади на основе кремния с учетом последовательного сопротивления $R_s$ следующее выражение

$$\eta = \frac{J_m V_m}{P_s}\left(1 - \frac{J_m R_s}{V_m}\right), \quad (6)$$

где $P_s$ =0.1 Вт/см$^2$ - плотность мощности падающего солнечного излучения при условиях освещения AM1.5G. Отметим, что при записи (4) значение температуры СЭ не конкретизовалось, а это означает, что полученные выражения для $V_m$ и $J_m$, а следовательно и выражение для $\eta$ будут справедливы при произвольной температуре.

Исходные параметры материала для HIT элемента и для двух образцов кремниевых СЭ с диффузионными p-n переходами, полученные при использовании результатов работ [1,2,6,9], приведены в таблице 1. С их использованием на рисунке 1 построены расчетные зависимости температурного коэффициента падения эффективности фотопреобразования с ростом температуры $K(T) = -\frac{d\eta(T)}{dT} / \eta(T)$. Отметим, что при расчете $K(T)$ предполагалось, что величины $J_{SC}$, $\tau_{SR}$ и $S$ не зависят от температуры. Основанием для этого является то, что основной вклад в зависимость $K(T)$ вносит температурная зависимость собственной концентрации носителей заряда в кремнии $n_i(T)$. Оценить изменение плотности тока короткого замыкания в кремнии при изменении температуры можно, учитывая уменьшение ширины запрещенной зоны кремния $E_g$ с увеличением температуры. Как показывают оценки, это приводит к росту $J_{SC}$ при изменении температуры от 298 до 340 К менее, чем на 1 %. Как показывают оценки, такого же порядка изменение эффективности фотопреобразования за счет учета температурной зависимости объемного времени жизни и скорости поверхностной рекомбинации.

Кривая 1 получена при использовании расчетных параметров для HIT элемента, приведенных в первой строчке таблицы 1. Как видно, в этом случае реализуется наименьшее значение величины $K(T)$, ~ 0.3 %/К. Это достигается за счет реализации наименьших значений $S$ (~ 9 см/с). Полученные расчетные данные хорошо коррелируют с экспериментальными значениями $K(T)$, приведенными для HIT элементов в [10]. Несколько большее значение $K(T)$ получено при использовании расчетных параметров для рекордного кремниевого СЭ на основе $p-n$ перехода, приведенных в второй строчке таблицы 1 (кривая 2). В данном случае величина $S$ равна 47 см/с. И, наконец, наибольшее значение $K(T)$ реализуется при использовании параметров кремниевого СЭ космического назначения, приведенных в третьей строчке таблицы 1. При $T$=300 К оно составляет около 0.45 %.К. В этом случае величина $S$= 700 см/с. Приведенное значение $K$ соответствует экспериментальному значению для коммерческих кремниевых СЭ на основе диффузионных p-n переходов, приведенному в работе [10].

Таким образом, как видно из рис. 1, величина $K$ коррелирует со значением $S$. Наименьшие значения $K$ реализуются для HIT элементов, в которых для получения минимальных значений $S$ используются возможности пассивации границ раздела, связанные с применением анизотипного и изотипного гетеропереходов на основе $\alpha - Si:H$. В то же время близкие значения $K$ получаются и для кремниевых СЭ на основе диффузионных p-n переходов, в которых величина скорости поверхностной рекомбинации минимизирована благодаря использованию термического окисла кремния [1, 11]. Необходимо отметить, что величина $K(T)$ непосредственно не определяется конструкцией СЭ, как это утверждается в [10].

## 2. Эффективность фотопреобразования с учетом конвекции и радиационного охлаждения для произвольных значений температуры окружающей среды

Вначале проанализируем температурные зависимости величин эффективностей фотопреобразования для HIT элемента, описанного в работе [2], а затем для рекордного СЭ с диффузионным p-n переходом (см. [1]).

### 2.1. Моделирование HIT элементов

Поскольку при получении выражения (6) температура СЭ не конкретизовалась, то мы можем определить эффективность СЭ и при произвольных значениях температуры окружающей среды $T_0$. В случае, когда решение получается самосогласованным образом, то вначале, подставив выражение (6) в (1), нужно решить полученное уравнение, определив величину $T$.

На рис. 2 приведены зависимости величины $T$ от коэффициента конвекции $\gamma$. При расчетах здесь и далее считалось, что $\beta$=2, а $K_T$=1, $E_1$ =1.12 эВ, $E_2$=10 эВ. В качестве $T_0$ были использованы следующие значения: 288, 298 и 308 K, что отвечает 15°, 25° и 35° C. Как видно из рисунка 1, во всем диапазоне изменения $\gamma$ ( от 10 до $3\cdot10^2$ Вт/м$^2$ K) температура СЭ остается больше, чем величины $T_0$. Величина $\gamma_0$ в работе [7] для различных условий эксперимента изменяется в пределах от 2.8 до 8.55 Вт/м$^2$ K, а величина $\delta$ - в пределах от 2.2 до 3.8 Вт·с/м$^3$ K. В работе [9] реперными значениями $\gamma$ являются величины 20 и 60 Вт/м$^2$ K. Первое значение описывает случай, когда скорость ветра мала по сравнению со значением 1 м/с, а второе отвечает усредненному значению скорости ветра, примерно равной 7 м/с. Хотя работы [7] и [12] дают несколько разные скорости ветра в реальных случаях, однако значение 300 Вт/м$^2$ K отвечает очень большой (ураганной) скорости ветра согласно данным, приведенным в них.

Отметим, что при приведенных выше значениях $\gamma$ конвекционный механизм охлаждения значительно эффективнее механизма радиационного охлаждения. Усилить радиационный механизм охлаждения можно за счет использования радиаторов, т.е. реализовав условие $K_T >> 1$.

На рис. 3 приведены самосогласованные по значению температуры зависимости величины $\eta$ от $\gamma$ для тех же величин $T_0$, что и на рис. 2. Помимо этого, на рис. 3 приведены значения $\eta$, реализующиеся тогда, когда температура СЭ равна температуре окружающей среды. Как видно из рисунка, реальные значения $\eta$ меньше, чем значения $\eta(T_0)$ даже при максимальных величинах $\gamma$, использованных при расчете. В частности, для реперной точки $\gamma$=60 Вт/м$^2$ K соответственно имеем: $\eta(288K)$=25.5 %, в то время как самосогласованное значение $\eta$ равно 24.7 %; $\eta(298K)$=24.7 %, а самосогласованное значение $\eta$ равно 24 %; $\eta(308K)$=24%, а самосогласованное значение $\eta$ равно 23.3 %. Относительное уменьшение величины $\eta$ из-за того, что охлаждение не происходит до температуры окружающей среды $T_0$, примерно равно 3 % для всех трех случаев.

### 2.2. Моделирование кремниевых СЭ с $p-n$ переходом и рекордной эффективностью фотопреобразования

На рис. 4 приведены зависимости величины $T$ для кремниевого p-n перехода от коэффициента конвекции $\gamma$. При этом в качестве $T_0$ были использованы такие же, как и раньше, значения: 288, 298 и 308 K. Как видно из рисунка 4, во всем диапазоне изменения $\gamma$ (от $10^{-3}$ до $3\cdot10^{-2}$ Вт/см$^2$ K) температура кремниевого СЭ больше, чем величины $T_0$.

На рис. 5 приведены самосогласованные по температуре зависимости величины $\eta$ от $\gamma$ для тех же значений $T_0$, что и на рис. 4. Помимо этого, на рис. 5 приведены значения $\eta$, реализующиеся тогда, когда температура СЭ равна температуре окружающей среды. Как видно из рисунка, как и для случая HIT элементов, реальные значения $\eta$ меньше, чем значения $\eta(T_0)$ даже при максимальных величинах $\gamma$, использованных при расчете. В частности, для реперной точки $\gamma$=60 Вт/м$^2$ К соответственно имеем: $\eta(288K)$=26.1 %, в то время как самосогласованное значение $\eta$ равно 25 %; $\eta(298K)$=25 %, а самосогласованное значение $\eta$ равно 24.2 %; $\eta(308K)$=24.5 %, а самосогласованное значение $\eta$ равно 23.3 %. Относительное уменьшение величины $\eta$ из-за того, что охлаждение не происходит до значений температуры окружающей среды $T_0$, в случае кремниевого p-n перехода соответственно равно 4.4 % при $T_0$=288 К, 3.4 % при $T_0$=298 К и 4.7 % при $T_0$=308 К.

В заключение отметим, что подобные отличия между реальными значениями эффективности фотопреобразования и значениями эффективности фотопреобразования при температуре, равной температуре окружающей среды, будут иметь место для всех кремниевых СЭ, работающих в земных условиях. Более того, как следует из уравнения (1), при $\eta \le 20\%$ указанные отличия будут выражены более сильно, чем для рассмотренных выше кремниевых СЭ. Это связано с тем, что чем меньше эффективность фотопреобразования, тем сильнее греется СЭ.

### 3. Случай концентрированного освещения

Как было сказано в постановке задачи, при реализации критерия $\Delta p \ge p_0$ величина напряжения разомкнутой цепи высокоэффективных кремниевых СЭ увеличивается за счет того, что заметный вклад в величину $V_{OC}$ дает тыльная поверхность. Ранее этот эффект обсуждался для случая, когда либо реализуются условия АМ1,5, либо интенсивность освещения и спектр поглощения отвечают этим же условиям. В то же время приведенное неравенство в высокоэффективных кремниевых СЭ, как правило, будет хорошо выполняться в условиях концентрированного освещения. В связи с этим проанализируем поведение $V_{OC}$ в зависимости от степени концентрации солнечного освещения $M$. Будем считать, что при этом ток короткого замыкания СЭ в М раз превышает ток короткого замыкания в условиях АМ1,5. Промоделируем зависимости $V_{OC}(M)$, основываясь, в основном, на параметрах рекордного кремниевого СЭ с эффективностью 25 %. На рис. 6 при использовании указанных параметров приведены зависимости $V_{OC}(M)$, построенные при использовании формулы (2) с учетом уравнения (3), из которого определялась величина $\Delta p$ (см. кривые 1-3). Параметром кривых является уровень легирования базы $p_0$, соответственно равный $10^{15}$, $10^{16}$ и $10^{17}$ см$^{-3}$.

Для сравнения приведены зависимости $V_{OC}(M)$, полученные при отбрасывании второго слагаемого в правой части уравнения (2). Они описывают обычный механизм зависимости $V_{OC}$ от уровня освещенности. Температура СЭ при расчете считалась равной 298 K, что отвечает принудительному охлаждению СЭ. Как видно из рис. 6, кривые 1-3 при высоких значениях $M$ совпадают. Это происходит, когда выполняется неравенство $\Delta p >> p_0$. Значения $V_{OC}$, полученные при использовании (2), существенно превышают значения, полученные при использовании усеченной формулы (особенно при больших уровнях освещенности). При освещенности 1000 Солнц достигается величина $V_{OC}$, равная 0.89 В, типичная скорее для более широкозонных полупроводников, чем для кремния.

Перейдем теперь к моделированию зависимостей эффективности фотопреобразования от степени концентрации солнечного освещения $M$ с учетом реальной температуры кремниевого СЭ. Будем считать, что температура СЭ постоянна во всех сечениях полупроводника. Это накладывает определенные ограничения на соотношение между теплопроводностью кремния и теплопроводностью граничащих с ним материалов. Для простоты также будем считать, что степень охлаждения СЭ как за счет радиационного, так и за счет конвекционного механизмов пропорциональна величине $M$. Практически это достигается за счет использования радиаторов, полная площадь которых растет пропорционально $M$. Тогда, учтя указанную зависимость в уравнении (1), и учтя также, что $J_{SC}(M) = MJ_{SC}(AM1.5)$, можем решить уравнения (1), (3) и (6) совместно, получив при этом искомую зависимость $\eta(M)$. Используем для расчета параметры рекордного СЭ с $p-n$ переходом. Результаты расчета приведены на рис. 7 (см. кривую 1). Для сравнения приведем также зависимость $\eta(M)$, полученную при предположении, что температура СЭ равна 298 K (см. кривую 2). Как видно из рисунка, при использованных для расчета параметрах рост $\eta(M)$ имеет место лишь при $M \le 10$. При $M > 10$ рост $\eta(M)$ заменяется спадом, что связано с влиянием последовательного сопротивления $R_s$, равного в рассматриваемом случае 0.15 Ом. Как известно, в СЭ, используемых при концентрированном освещении, величина $R_s$ существенно меньше, чем в СЭ, работающих при неконцентрированном освещении. Это учтено при построении зависимости $\eta(M)$ для реальной температуры СЭ при использовании значения $R_s = 0.01$ Ом (см. рис. 7, кривая 2). Кривая 2' соответствует случаю, когда $R_s = 0.01$ Ом, а температура СЭ равна 298 K. В данном случае зависимости $\eta(M)$ растущие вплоть до значения $M \approx$ 100, затем величина $\eta$ проходит через максимум и начинается спад. Как и в предыдущем случае, спад зависимостей $\eta(M)$ связан с влиянием последовательного сопротивления. Из рис. 7 видно, что величина $\eta(M)$ при реальной температуре меньше, чем при температуре окру-жающей среды, равной 298 К. Как и в разделе 2 это связано с тем, что реальная температура СЭ больше температуры окружающей среды.

### 4. Заключение

Предложенный в данной работе подход позволяет рассчитать реальные значения эффективности фотопреобразования как для высокоэффективных HIT элементов, так и для СЭ с кремниевым p-n переходом. Как показал самосогласованный расчет эффективности фотопребразования СЭ с учетом реальной температуры, величина $\eta$ занижена по сравнению со значениями, получаемыми, например, в условиях AM 1.5, причем ее занижение в случае конкретных параметров СЭ, соответствующих приведенным в работе [3], равно от 3 до 5 %. Следует отметить, что подобные отличия будут иметь место для всех кремниевых СЭ, работающих в земных условиях, а не только для высокоэффективных.

Показано, что при концентрированных уровнях освещенности в высокоэффективных кремниевых СЭ величина напряжения разомкнутой цепи дополнительно возрастает из-за вклада в величину $V_{OC}$ фотоэдс, возникающей на тыльной поверхности. Этот эффект весьма существенный. Так, прирост величины $V_{OC}$ в случае, когда M= 1000, а $p_0 = 10^{15}$ см$^{-3}$, составляет 0.14 В. Определены расчетные зависимости эффективности фотопреобразования высокоэффективных кремниевых СЭ от степени концентрации освещения при реальных температурах СЭ для эффективно действующих механизмов охлаждения.

Литература

Подписи к рисункам

Рис. 1. Расчетные зависимости температурного коэффициента спада эффективности фотопреобразования $K$ от температуры для HIT элемента (кривая 1), рекордного СЭ с p-n переходом (кривая 2) и СЭ с p-n переходом космического назначения (кривая 3). Значения, приведенные на кривых 4 и 5, взяты из [10].

Рис. 2. Зависимости температуры HIT элемента от коэффициента конвекции $\gamma$. При построении рисунка принято, что температура окружающей среды соответственно равна 15, 25 и 35° С (кривые 1' -3').

Рис.3. Зависимости эффективности фотопреобразования HIT элемента от коэффициента конвекции $\gamma$. Рис. 2a, 2b, 2c отвечают температуре окружающей среды, равной 15, 25 и 35 ° С.

Рис. 4. Зависимости температуры кремниевого p-n перехода от коэффициента конвекции $\gamma$. При построении рисунка принято, что температура окружающей среды соответственно равна 15, 25 и 35° С (кривые 4 -6).

Рис. 5. Зависимости эффективности фотопреобразования кремниевого p-n перехода от коэффициента конвекции $\gamma$. Рис. 4a, 4b, 4c отвечают температуре окружающей среды, соответственно равной 15, 25 и 35 ° С.

Рис. 6. Зависимости напряжения разомкнутой цепи от степени концентрации солнечного освещения. Кривые 1-3 построены с использованием полной формулы (2), а кривые 1'-3' – с использованием усеченной формулы для уровней легирования полупроводника $10^{15}$, $10^{16}$ и $10^{17}$ $см^{-3}$ соответственно.

Рис.7. Зависимости эффективности фотопреобразования от степени концентрации солнечного освещения. Кривые 1 и 2 построены для случая реальной температуры СЭ, а кривые 1'-2' – для температуры СЭ, равной 298 К. Использованные параметры: $R_s$, Ом: 1, 1' – 0.15; 2, 2' – 0.01.

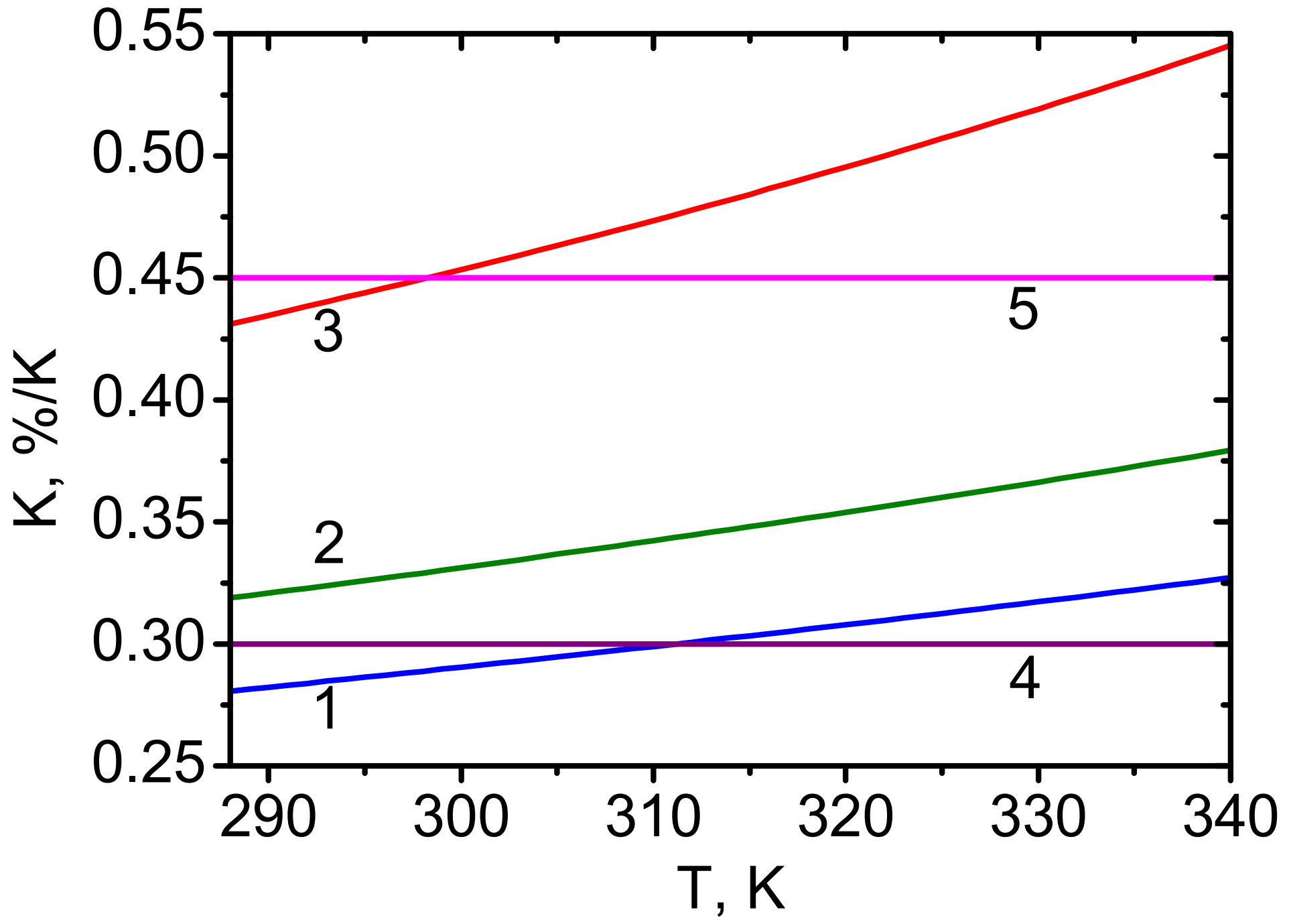

Рис. 1 к статье А.В. Саченко и др. ”Моделирование реальной эффективности…”

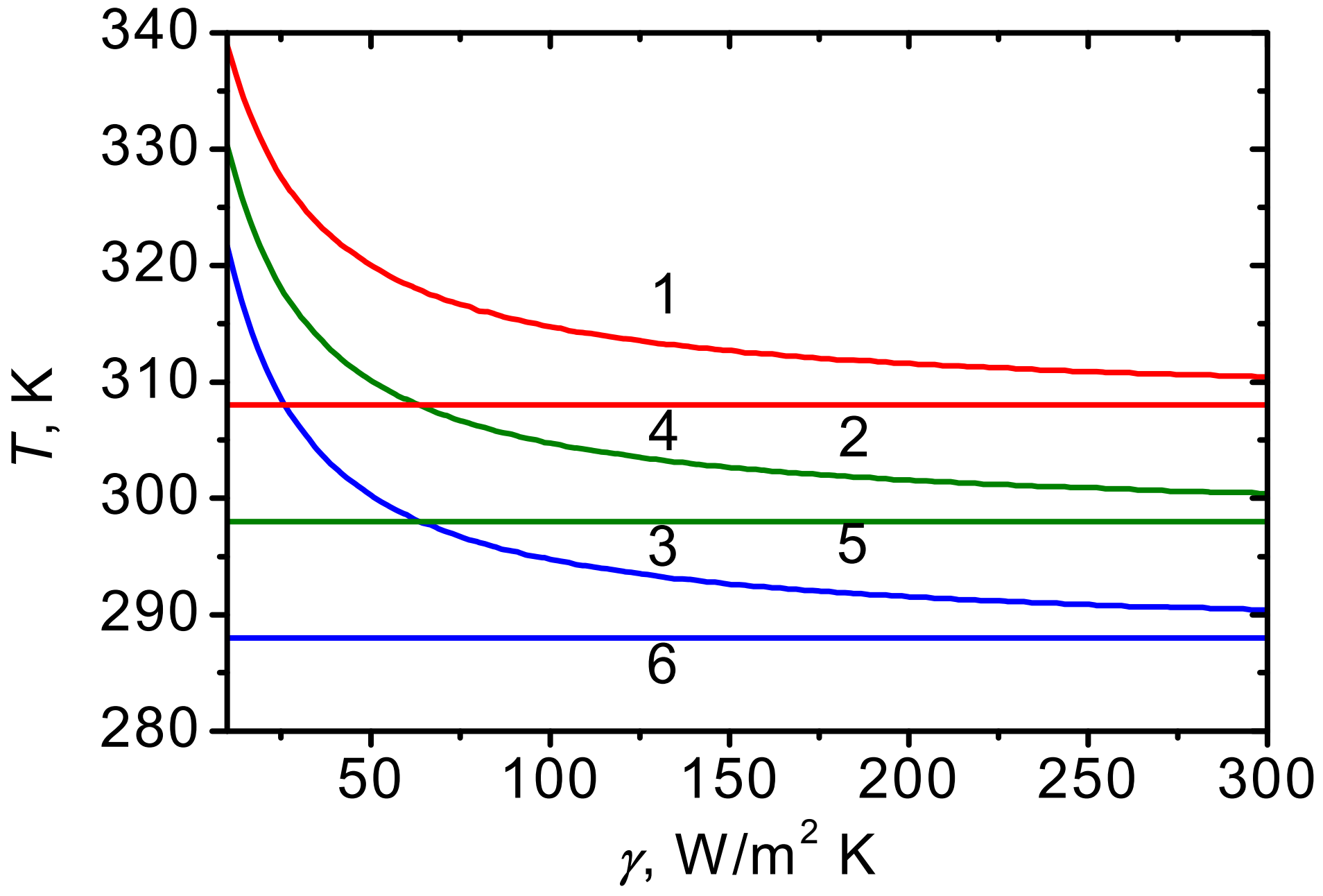

Рис. 2 к статье А.В. Саченко и др. ”Моделирование реальной эффективности…”

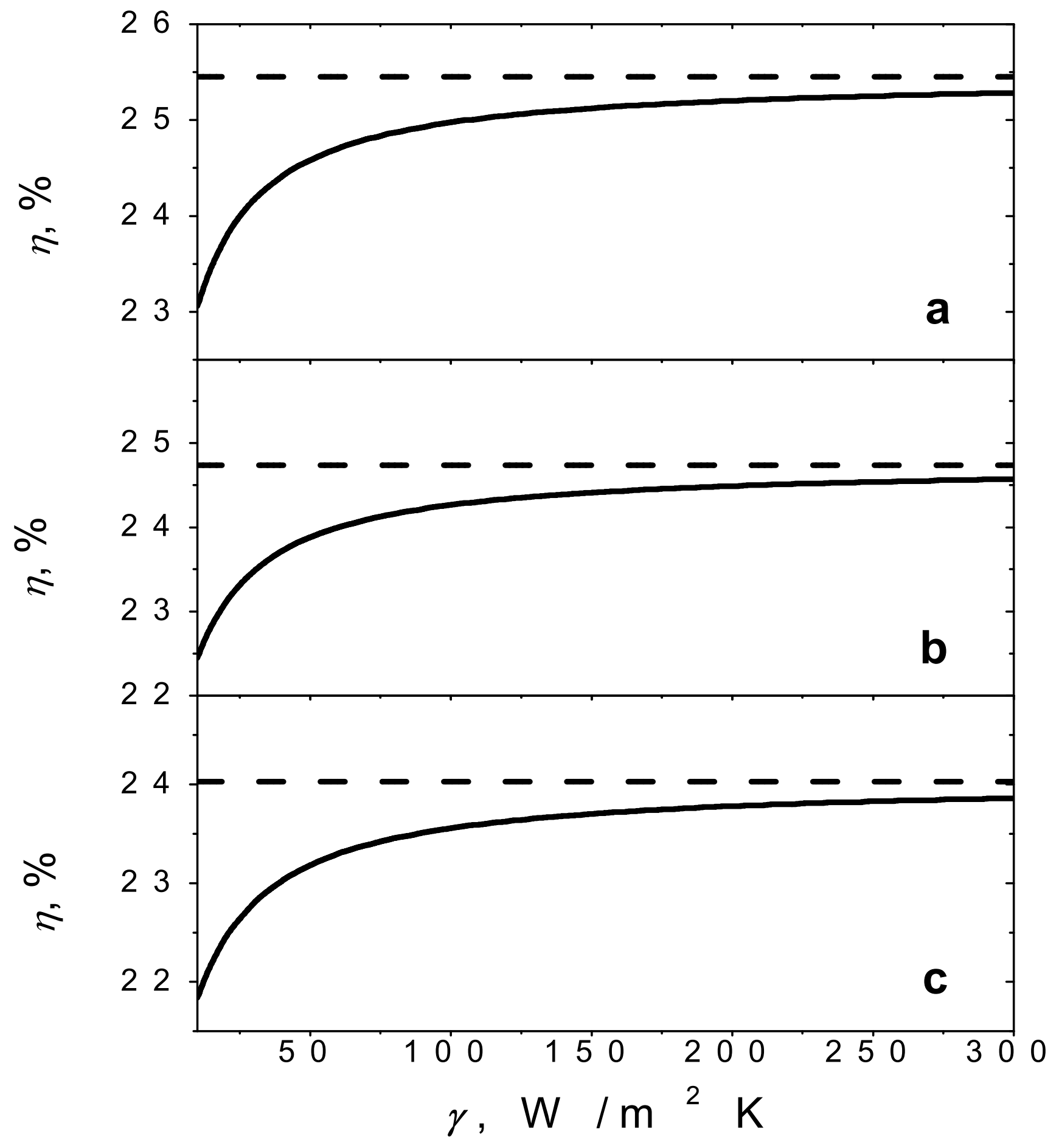

Рис. 3 к статье А.В. Саченко и др. ”Моделирование реальной эффективности…”

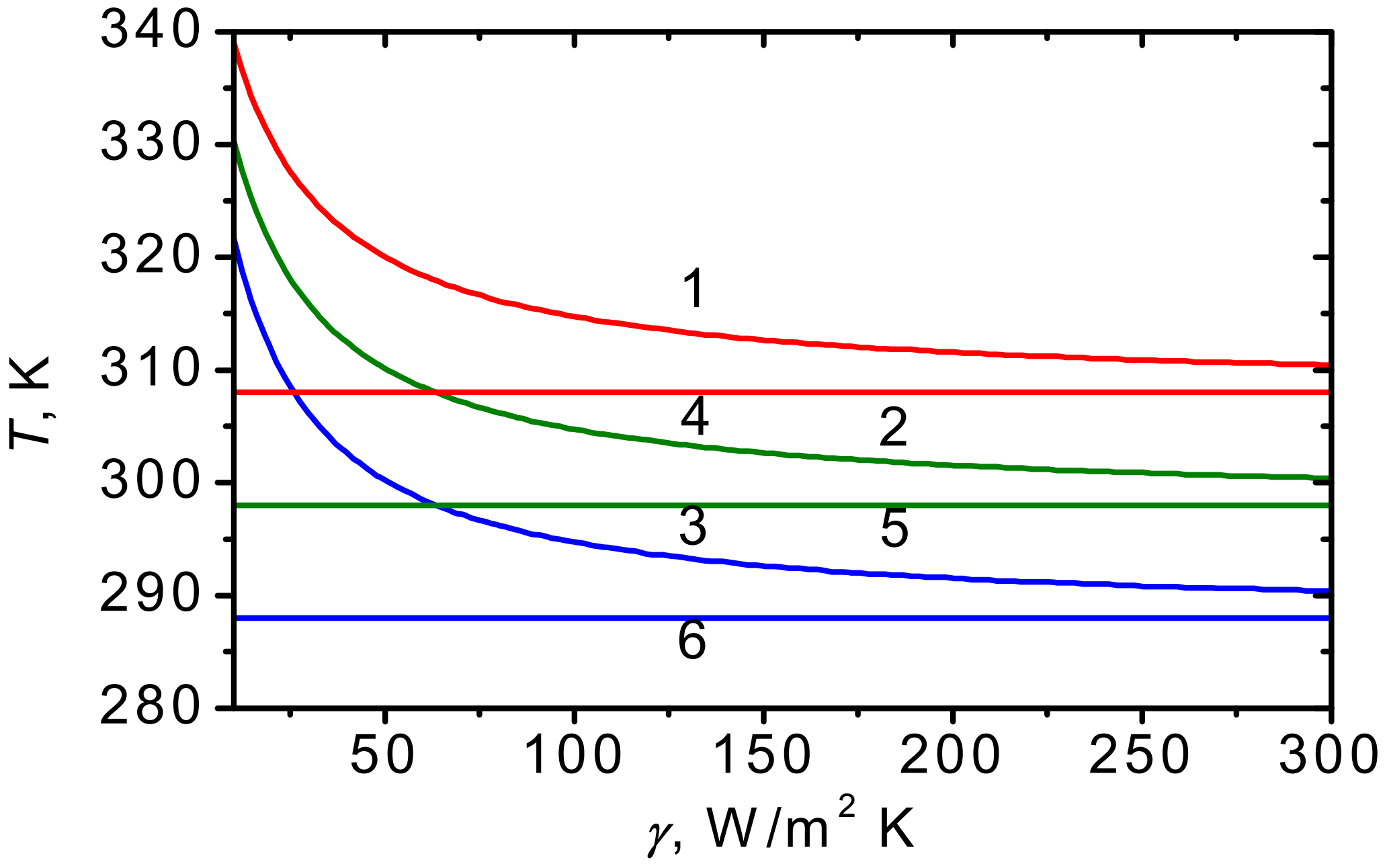

Рис. 4 к статье А.В. Саченко и др. ”Моделирование реальной эффективности…”

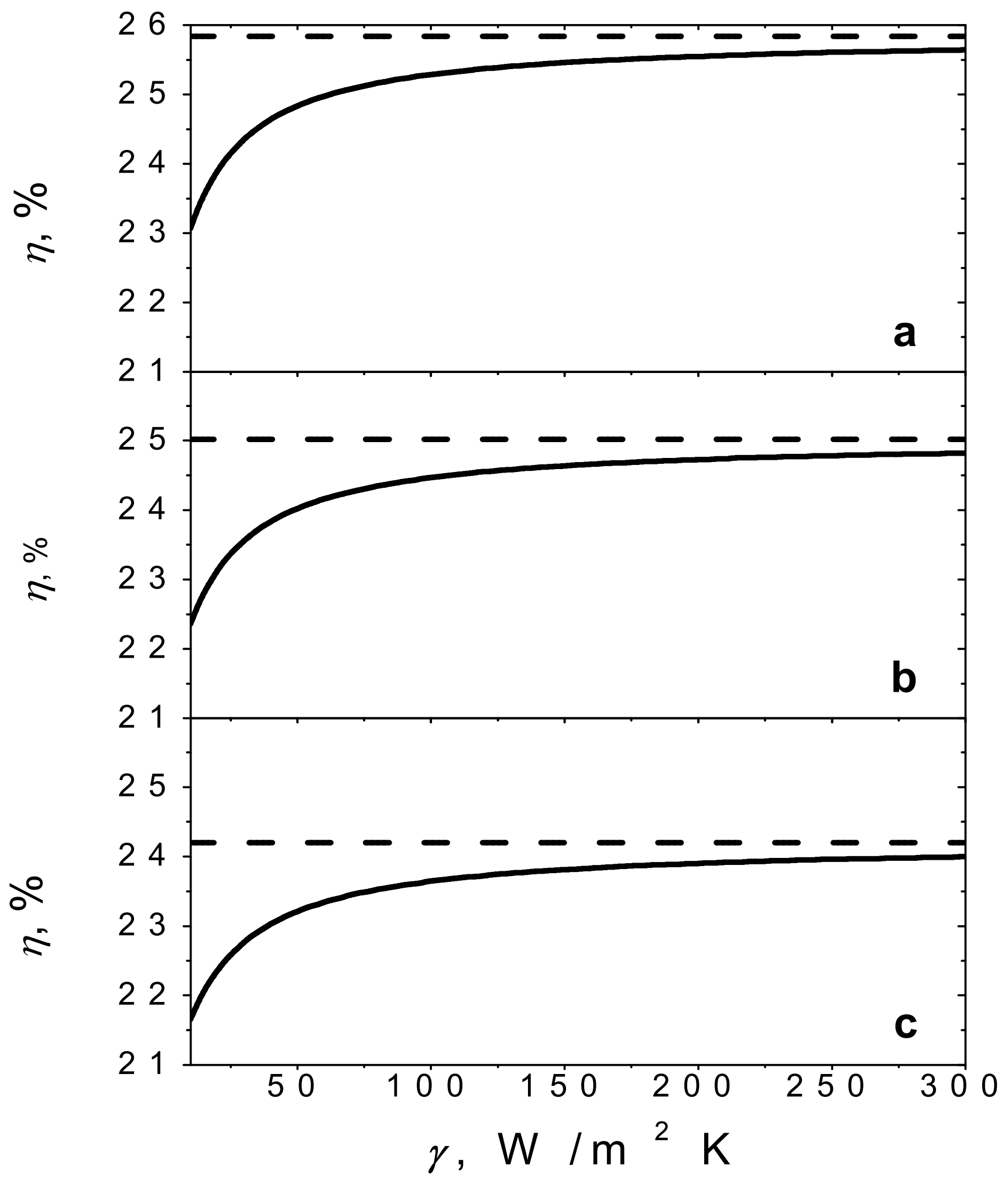

Рис. 5 к статье А.В. Саченко и др. ”Моделирование реальной эффективности…”

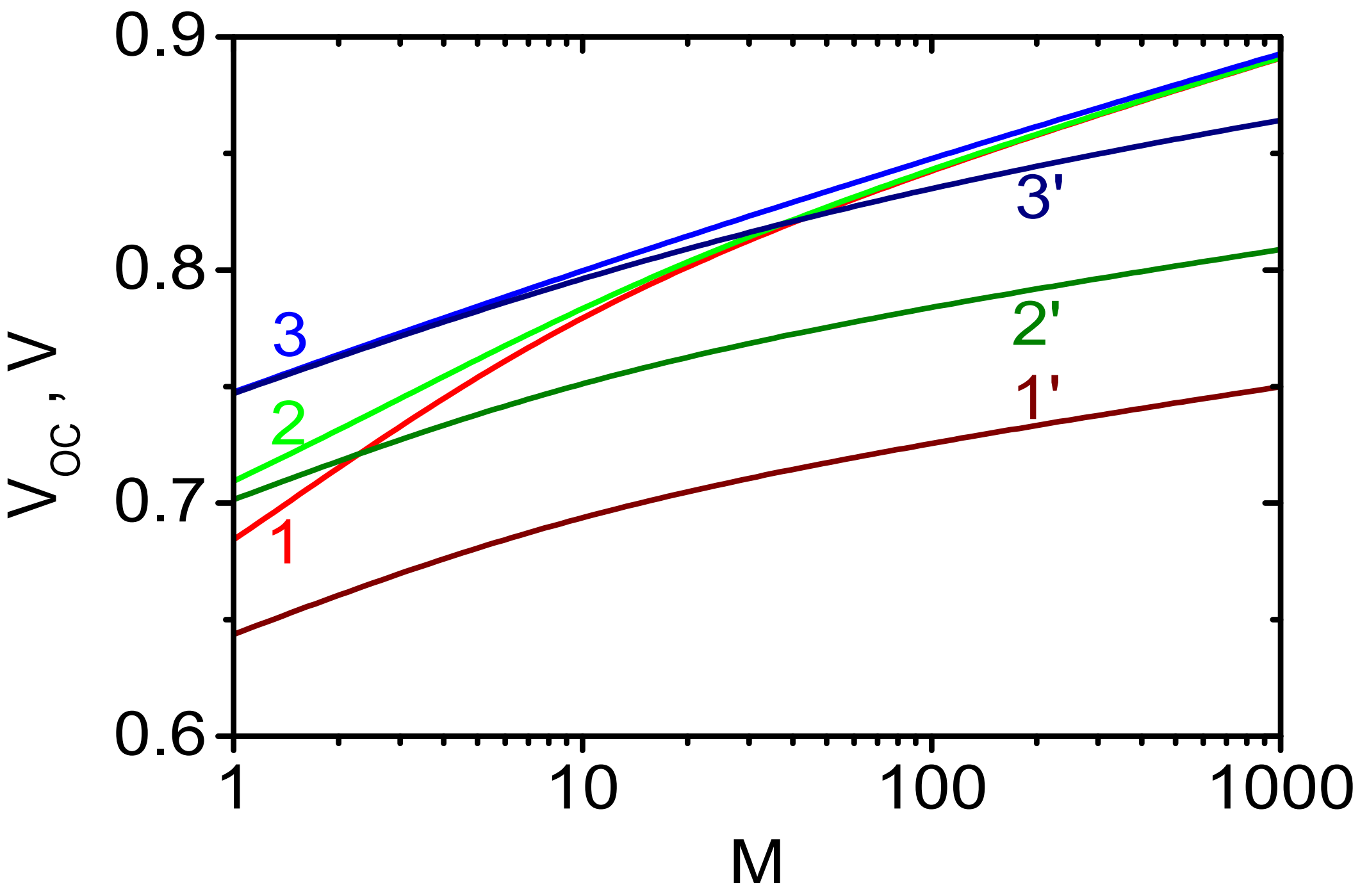

Рис. 6 к статье А.В. Саченко и др. ”Моделирование реальной эффективности…”

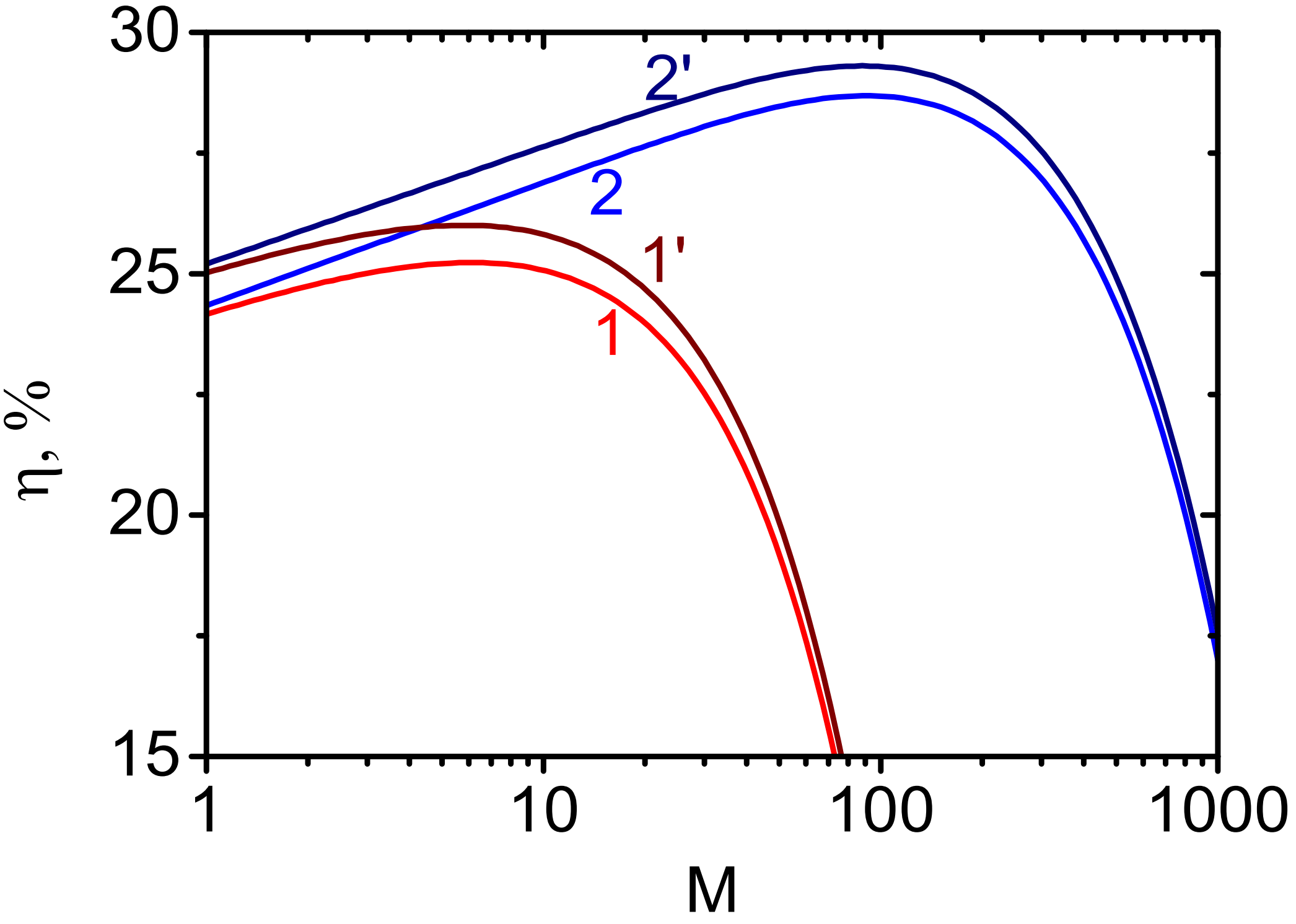

Рис. 7 к статье А.В. Саченко и др. ”Моделирование реальной эффективности…”

Таблица 1

| Number cell | $J_{SC}^{\exp}$, mA/cm$^2$ | d, $\mu$m | S, cm/s | $\tau_{SR}$,ms | $n_0$, cm$^{-3}$ | $R_S$,Ohm |
|---|---|---|---|---|---|---|
| 1 | 39.5 | 100 | 4 | 1 | $8\cdot10^{15}$ | 0.04 |
| 2 | 42.7 | 200 | 47 | 0.87 | $9.3\cdot10^{15}$ | 0.15 |
| 3 | 35.0 | 380 | 700 | 0.5 | $3.1\cdot10^{15}$ | 0.55 |